\documentclass[a4paper, fleqn, usenatbib]{mnras}
\usepackage{ulem}       
\usepackage[T1]{fontenc}
\usepackage{ae,aecompl}
\usepackage{graphicx}   
\usepackage{epstopdf}	
\usepackage{amsmath}    
\usepackage{amssymb}    
\usepackage{times}
\usepackage{txfonts}
\usepackage{siunitx}            

\title[Radiative transfer in disc galaxies V]{Radiative transfer in disc galaxies --  V. The accuracy of the KB approximation}

\author[D. Lee et al.]{%
Dukhang Lee,$^{1,2,3}$\thanks{E-mail: lee.dukhang@gmail.com}
Maarten Baes,$^{1}$
Kwang-Il Seon,$^{2,3,4}$
Peter Camps,$^{1}$
Sam Verstocken$^{1}$
\newauthor and
Wonyong Han$^{2,3}$
\\
$^{1}$Sterrenkundig Observatorium, Universiteit Gent, Krijgslaan 281 S9,
B-9000 Gent, Belgium \\
$^{2}$Korea Astronomy and Space Science Institute, Daejeon, 305-348 Korea\\
$^{3}$Astronomy and Space Science Major, University of Science and Technology, Daejeon 305-350, Korea\\
$^{4}$Department of Astrophysical Sciences, Princeton University, Princeton, NJ 08544, USA
}

\date{Accepted XXX. Received YYY; in original form ZZZ}

\pubyear{2016}
\usepackage{color}

\begin{document}

\label{firstpage}
\pagerange{\pageref{firstpage}--\pageref{lastpage}}
\maketitle

\begin{abstract}
We investigate the accuracy of an approximate radiative transfer technique that was first proposed by Kylafis \& Bahcall (hereafter the KB approximation) and has been popular in modelling dusty late-type galaxies. We compare realistic galaxy models calculated with the KB approximation with those of a three-dimensional Monte Carlo radiative transfer code \textsc{skirt}. The  \textsc{skirt} code fully takes into account of the contribution of multiple scattering whereas the KB approximation calculates only single scattered intensity  and multiple scattering components are approximated. We find that the KB approximation gives fairly accurate results if optically thin, face-on galaxies are considered. However, for highly inclined ($i \gtrsim 85\degr$) and/or optically thick (central face-on optical depth
$\gtrsim1$) galaxy models, the approximation can give rise to substantial errors, sometimes, up to $\gtrsim 40\%$. Moreover, it is also found that the KB approximation is not always physical, sometimes producing infinite intensities at lines of sight with high optical depth in edge-on galaxy models. There is no ``simple recipe'' to correct the errors of the KB approximation that is universally applicable to any galaxy models. Therefore, it is recommended that the full radiative transfer calculation be used, even though it's slower than the KB approximation. 
\end{abstract}



\section{Introduction}

The presence of interstellar dust in galaxies severely affects the way in which we observe galaxies. Interstellar dust grains efficiently scatter and absorb ultraviolet (UV), optical and near-infrared radiation, and reradiate the absorbed energy at infrared and submm wavelengths. Over the past decades, it has been demonstrated repeatedly that the effects of dust radiative transfer in galaxies are complex and often counter-intuitive \citep{1989MNRAS.239..939D, 1994ApJ...432..114B, 2001MNRAS.326..733B, 2004A&A...419..821T, 2010MNRAS.403.2053G}, and hence that full dust radiative transfer modelling is required in order to fully interpret the observed properties of galaxies.

Dust radiative transfer modelling is complex, since the specific intensity (the quantity that needs to be solved) depends on space, propagation direction, wavelength and time. Moreover, the radiative transfer equation is a complex integro-differential equation that is both nonlinear and nonlocal. As a result, solving the radiative transfer problem in a general three-dimensional (3D) context is demanding and time-consuming. It is hence useful to investigate whether this complex problem can be simplified in some way. 

An interesting and original approach has been proposed by \citet{1987ApJ...317..637K}, hereafter referred to as the KB approximation. Inspired by early radiative transfer work by \citet{1937ApJ....85..107H} and \citet{1969Phy....41..151V}, \citet{1987ApJ...317..637K} wrote the specific intensity of the radiation field as a series of terms, in which each term represents the contribution to the radiation field of radiation that has been scattered exactly $n$ times. They argued that this series fairly quickly shows a geometrical behaviour. This allows estimating the higher-order terms, and hence the total intensity, based on the first-order terms only.  \citet{1987ApJ...317..637K} first used this approach to model the dust distribution in the edge-on spiral galaxy NGC\,891, and the approach was extended and used later on to model several other edge-on spiral galaxies \citep{1997A&A...325..135X, 1999A&A...344..868X}. The KB approximation was used quite extensively in the following years in different radiative transfer problems, for example to quantify the attenuation signatures in disc galaxies \citep{1994ApJ...432..114B, 2004A&A...419..821T, 2013A&A...553A..80P}, to investigate the effects of spiral structure on dust lanes in edge-on spirals \citep{2000A&A...353..117M}, to derive correction factors for the change in the apparent disc scalelengths and central surface brightness due to the effect of dust in disc galaxies \citep{2006A&A...456..941M}, and to model the spectral energy distributions of dusty galaxies \citep{2000A&A...362..138P, 2011A&A...527A.109P, 2001A&A...372..775M}.

In the past few years, the field of dust radiative transfer has changed drastically, thanks to a combination of increased computing power and the development of new algorithms and acceleration techniques. It is now possible to solve the full dust radiative transfer problem (i.e. including absorption, scattering and thermal emission) in an arbitrary 3D geometry. Several sophisticated codes have been developed for this goal \citep[e.g.,][]{2001ApJ...551..269G, 2006A&A...459..797P, 2006MNRAS.372....2J, 2008A&A...490..461B, 2011A&A...536A..79R, 2012A&A...544A..52L, 2014MNRAS.438.3137N, 2015A&C.....9...20C, 2016arXiv160602030S}. The vast majority of these codes are based on the Monte Carlo technique \citep[see][for an overview]{2011BASI...39..101W}. 

Thanks to these new developments, we can now critically assess the validity, strengths and limitations of the KB approximation. This is the goal of the present paper. Based on the Monte Carlo radiative transfer code \textsc{skirt}\footnote{http://www.skirt.ugent.be} \citep{2003MNRAS.343.1081B, 2011ApJS..196...22B, 2015A&C.....9...20C}, we will compute mock observed images for a simple but realistic dusty disc galaxy with and without the KB approximation, and compare the results as a function of various input parameters. In Section 2, we describe the methods with and without the approximation, and we present the galaxy models used in the modelling. The results of the calculations are presented in Section 3. In Section 4 we discuss the accuracy and the applicability of the approximate radiative transfer algorithm. A summary is given in Section 5.


\section{Radiative transfer modelling}
\label{Modelling.sec}

\subsection{SKIRT radiative transfer}

\textsc{skirt} is a publicly available state-of-the-art dust radiative transfer code. As most 3D radiative transfer codes, \textsc{skirt} is based on the Monte Carlo technique. The essence of Monte Carlo radiative transfer is that the radiation field is treated as the flow of a large number of photon packages, whereby each individual photon package is followed along its journey through the dusty medium. At every stage in its journey, the characteristics that determine the path of a photon package are determined probabilistically by generating random numbers from an appropriate probability density function. A comprehensive description of Monte Carlo dust radiative transfer can be found in e.g.\ \citet{2011BASI...39..101W} or \citet{2013ARA&A..51...63S}. Special features of \textsc{skirt} are an extensive suite of input models \citep{2015A&C....12...33B}, a carefully designed interface \citep{2015A&C.....9...20C}, the use of advanced spatial grids to partition the dusty medium \citep{2013A&A...560A..35C, 2013A&A...554A..10S, 2014A&A...561A..77S}, and a number of novel Monte Carlo acceleration techniques \citep[e.g.,][]{2011ApJS..196...22B, 2016A&A...590A..55B}.

\textsc{skirt} has mainly been developed to model various types of galaxies \citep{2010A&A...518L..39B, 2010A&A...518L..54D, 2014A&A...571A..69D, 2010MNRAS.403.2053G, 2010A&A...518L..45G, 2016arXiv160506239M}, but it has also been applied to dusty tori around active galactic nuclei \citep{2012MNRAS.420.2756S, 2016MNRAS.458.2288S}, molecular clouds \citep{2015A&A...575A.110H}, pinwheel nebulae \citep{2016MNRAS.tmp..943H} and circumbinary discs \citep{2015A&A...577A..55D}.

\subsection{The KB approximation}
\label{KB.sec} 

As most other Monte Carlo radiative transfer codes, \textsc{skirt} calculates simulated images (and spectral energy distributions) by peeling off photon packages after every emission and scattering event \citep[for details, see][]{1984ApJ...278..186Y, 2008MNRAS.391..617B}, and gradually summing the contribution of each peel-off photon in every pixel of a detector array. As every photon package in the simulation stores the number of scattering events it has experienced in its life cycle, separate images can in principle be constructed corresponding to every individual scattering level. Denoting the intensity corresponding to radiation that has escaped from the model in the direction of the observer after being scattered exactly $n$ times as $I_n$ along a certain line of sight, the total intensity $I$ is given by
\begin{equation}
I=\sum_{n=0}^{\infty} I_n.
\label{I}
\end{equation}
\citet{1987ApJ...317..637K} introduced an approximation that avoids the need to calculate the higher order scattered intensities. They rewrite the series~(\ref{I}) as
\begin{equation}
\label{Irewritten}
I = I_0
\left[
1 
+ \left(\frac{I_1}{I_0}\right) 
+ \left(\frac{I_1}{I_0}\right)\left(\frac{I_2}{I_1}\right) 
+ \left(\frac{I_1}{I_0}\right)\left(\frac{I_2}{I_1}\right)\left(\frac{I_3}{I_2}\right)
+ \cdots
\right] 
\end{equation}
and then make the approximation that the ratios of consequent terms for all $n~(>1)$ are equivalent to the first ratio term ($I_1/I_0$),
\begin{equation}
\label{assumption}
\frac{I_{n+1}}{I_n} \approx \frac{I_1}{I_0}.
\end{equation}
Using the assumption~(\ref{assumption}), the surface brightness of a galaxy can be written as
\begin{equation}
\label{Iwithassumption}
I_{\text{KB}} 
= I_0\left[1 + \left(\frac{I_1}{I_0}\right) + \left(\frac{I_1}{I_0}\right)^2 + \cdots\right].
\end{equation}
If $I_1/I_0<1$, then equation~(\ref{Iwithassumption}) can be further simplified to 
\begin{equation}
I_{\text{KB}} = I_0\left(\frac{1}{1-\frac{I_1}{I_0}}\right),
\label{Iapprox}
\end{equation}
since the terms in the brackets in equation~(\ref{Iwithassumption}) is a geometric series with a constant ratio of $I_1/I_0$. As shown in equation~(\ref{Iapprox}), only two lowest terms $I_0$ and $I_1$ are required to calculate the surface brightness in the KB approximation, indicating the approximation method is much simpler and faster than the full radiative transfer which takes into account all higher-order scattering terms. 

On the other hand, it is also immediately clear that the KB approximation has its limitations. It can give infinite intensities when $I_1>I_0$ along some lines of sight (or negative values if equation (\ref{Iapprox}) is used instead of the general expression (\ref{Iwithassumption})). A particular case of this situation is when there is no stellar source along a certain line of sight, because in this case $I_0=0$. In our galaxy models, characterised by continuous distributions of stars and dust (see Section~{\ref{GalaxyModel.sec}}), this does not occur, but it is relevant for other common radiative transfer cases, like circumstellar discs or dust tori around active galactic nuclei.

In the present study, we used the \textsc{skirt} code to produce images of simulated spiral galaxies, in two different ways. First, we calculated images in the ``standard'' way, i.e. calculating the full series of partial intensities $I_n$. Additionally, we also computed images using the KB approximation (\ref{Iapprox}), only using the partial intensities $I_0$ and $I_1$. By doing so, potential effects of using different computation techniques in calculating $I_0$ and $I_1$ can be avoided. Consequently, any difference between the full radiative transfer model and the KB approximation would merely originate from the adoption of the approximation~(\ref{assumption}).

\subsection{Galaxy model}
\label{GalaxyModel.sec}

To compare the full radiative transfer calculation and the KB approximation, we constructed a simple but realistic galaxy model. The model is based on the galaxy models that have been proven to reproduce a realistic geometry of dust and stars in edge-on galaxies \citep[e.g.][]{1987ApJ...317..637K, 1997A&A...325..135X, 1999A&A...344..868X, 2007A&A...471..765B, 2010A&A...518L..39B, 2012MNRAS.427.2797D, 2014MNRAS.441..869D}. 

Our model consists of three components: an exponential stellar disc, a S\'ersic bulge and an exponential dust disc. For a complete description of the model and the meaning of the various parameters, we refer to Section~3.1 of \citet{2013A&A...550A..74D}. The parameter values are listed in Table~{\ref{ParametersTable}}, and correspond to the average $g$-band values of a set of 10 edge-on CALIFA spiral galaxies modelled by \citet{2014MNRAS.441..869D}. The maximum radial and vertical distances in the dust grid system are assumed to be 60 kpc and $\pm$20 kpc, respectively. The only free parameter in our model is the dust mass, or correspondingly, the central face-on optical depth $\tau$. We calculated the radiative transfer models with various central face-on optical depth to examine the accuracy of the KB approximation depending on the optical depth. 

\begin{table}
\centering
\caption{Galaxy model parameters for the stellar and dust components.}
\label{ParametersTable}
\begin{tabular}{lcc}
        \hline
        \vspace{-2.3 mm}        
                                                                &                                                               &       \\
        Component               & Parameter                             & Value \\ 
        \vspace{-2.3 mm}
                                                                &                                                               &       \\
        \hline
        \vspace{-2.3 mm}
                                                                &                                                               &       \\
        Stellar disc            & $h_{R}$ (kpc)                 & 4.23  \\
                                                                & $h_{z}$ (kpc)                   & 0.51  \\
                                                                & $L/L_{\text{tot}}$    & 0.59    \\
        \vspace{-2.1 mm}
                                                                &                                                               &       \\
        Bulge                           & $q$                                           & 0.56    \\
                                                                & $n$                                           & 2.61    \\
                                                                & $r_{\text{eff}}$ (kpc)   & 2.31  \\
                                                                & $L/L_{\text{tot}}$    & 0.41    \\
        \vspace{-2.1 mm}
                                                                &                                                               &       \\
        Dust disc                       & $h_{R}$ (kpc)                 & 6.03    \\
                                                                & $h_{z}$ (kpc)                   & 0.23  \\ 
        \vspace{-2.3 mm}
                                                                &                                                               &       \\
        \hline
\end{tabular}
\end{table}

\section{Results}
\label{Results.sec}

\begin{figure*}
\centering
\includegraphics[width=0.95\textwidth, trim={0 0.6cm 0 0}, clip]{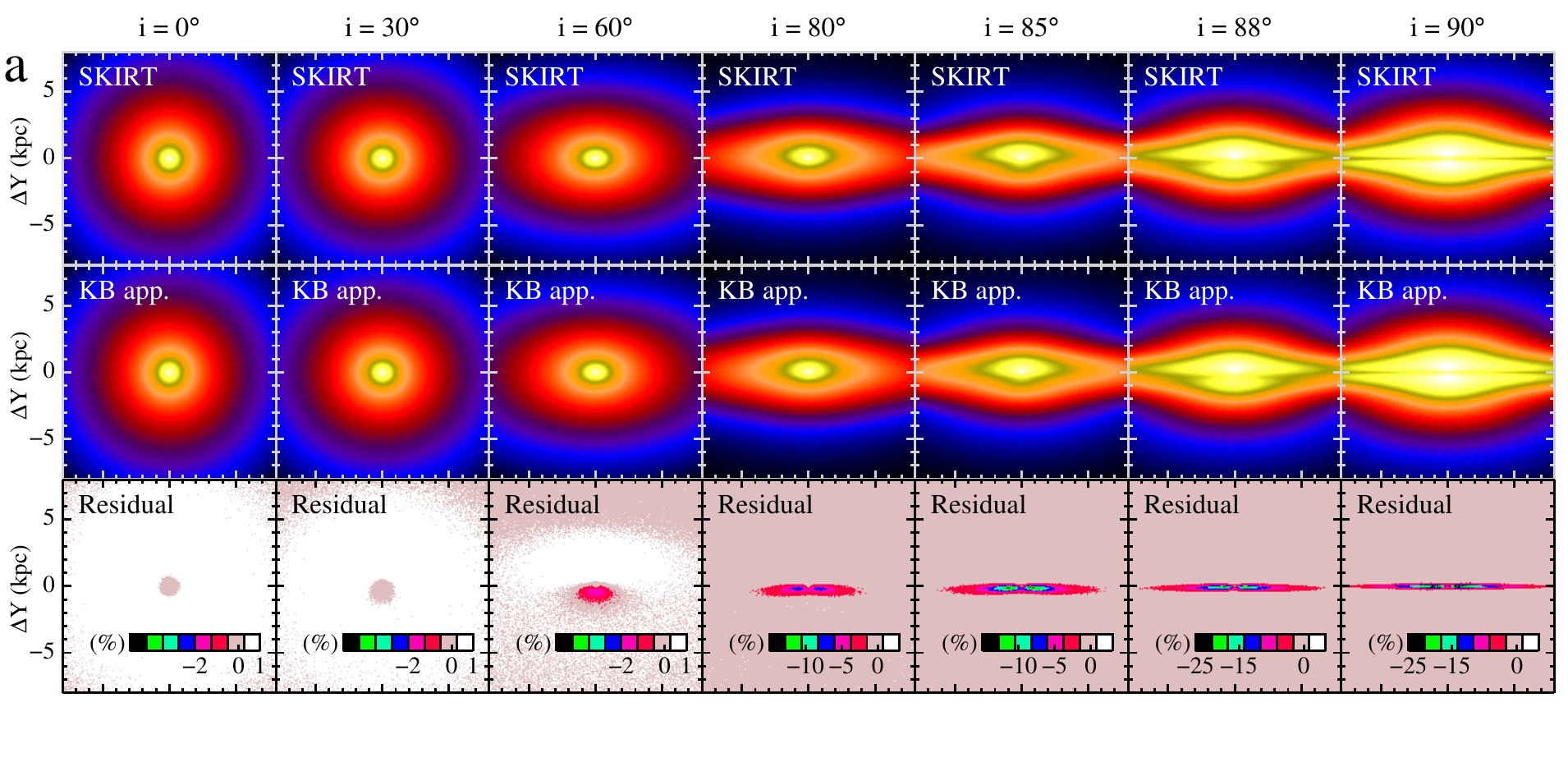}
\includegraphics[width=0.95\textwidth, trim={0 0.6cm 0 0.5cm}, clip]{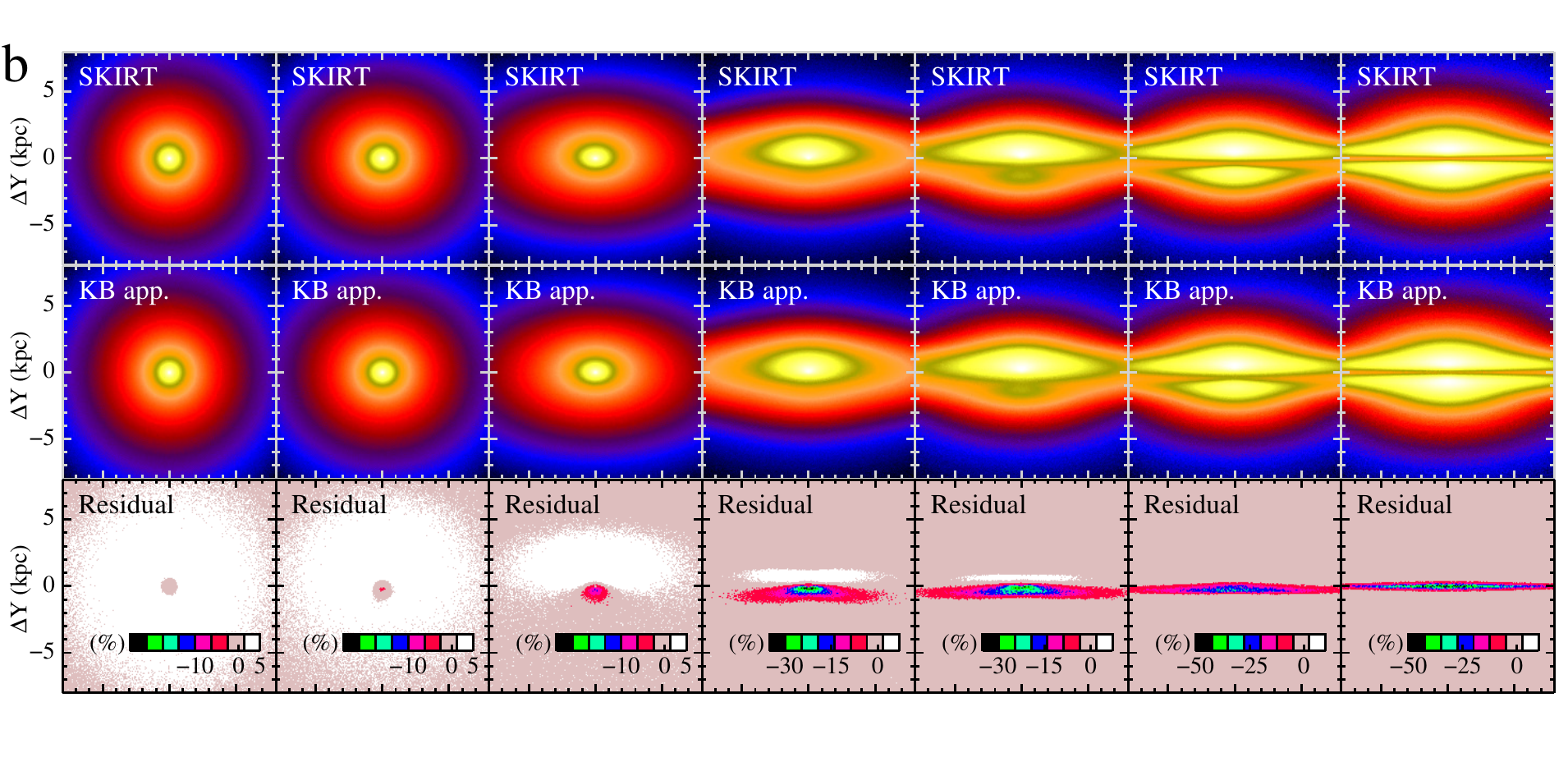}
\includegraphics[width=0.95\textwidth, trim={0 0 0 0.5cm}, clip]{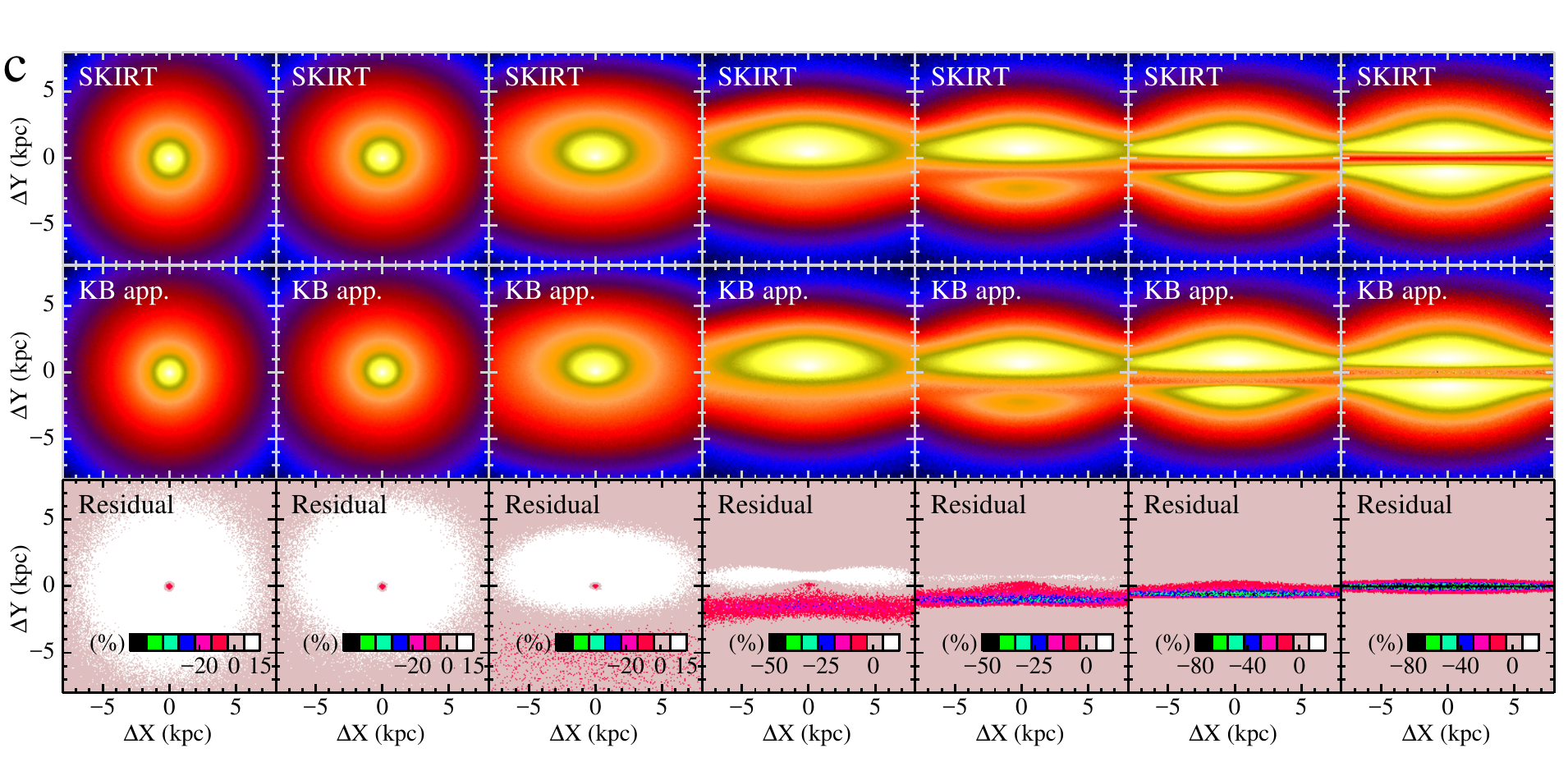}
\caption{Result images of the \textsc{skirt} code (top row), which is full radiative transfer without using the KB approximation, and the KB approximation (middle row) for $\tau_{\text{V}} =$ (a) 0.36, (b) 1.07 and (c) 3.56. The calculations were performed in the optical V band. Inclination angles are 0$\degr$, 30$\degr$, 60$\degr$, 80$\degr$, 85$\degr$, 88$\degr$ and 90$\degr$ from left to right. The bottom row shows error of the KB approximation models, defined by equation (\ref{residual}).}
\label{ModelResults.fig}
\end{figure*}

The main results of our calculations are shown in Fig.~{\ref{ModelResults.fig}}. This composite figure contains three blocks, corresponding to models with different levels of dust content. Within each block, model results are shown for the galaxy seen at different inclinations, ranging from completely face-on to completely edge-on. The top panels in each block show the images obtained with \textsc{skirt} without approximation, i.e. using formula~(\ref{I}). The panels on the second row show the corresponding images obtained using the KB approximation formula~(\ref{Iapprox}). All images in the first and second rows are shown on a logarithmic scale. 

Qualitatively, the two sets of images corresponding to a single model agree very well, which might suggest that the KB approximation works fairly well. More insight can be gained by looking at the residuals defined by
\begin{equation}
\label{residual}
{\cal{R}} = 100 \times \frac{I - I_{\text{KB}}}{I}, 
\end{equation}
where $I$ and $I_{\rm KB}$ are the results without and with the KB approximation, respectively. The bottom panels in each block of Fig.~{\ref{ModelResults.fig}} show the residual images on a linear scale. In these residual images, one can clearly notice that the approximation is not always accurate. 

In general, the KB approximation becomes less accurate as the optical depth and/or inclination angle increase. We also note that there are ``regional'' variations in the error: in some parts of the image the surface brightness is overestimated, whereas it is underestimated in other parts. In particular, the images typically show positive residuals in the upper half of the disc (where the stellar light suffers less extinction) and negative residuals in the lower half, especially at high inclinations ($60\degr \leq i \leq 85\degr$). The residuals in the bottom half are in general broader and have larger absolute values. 

\begin{figure*}
\centering
\includegraphics[width=0.5\textwidth]{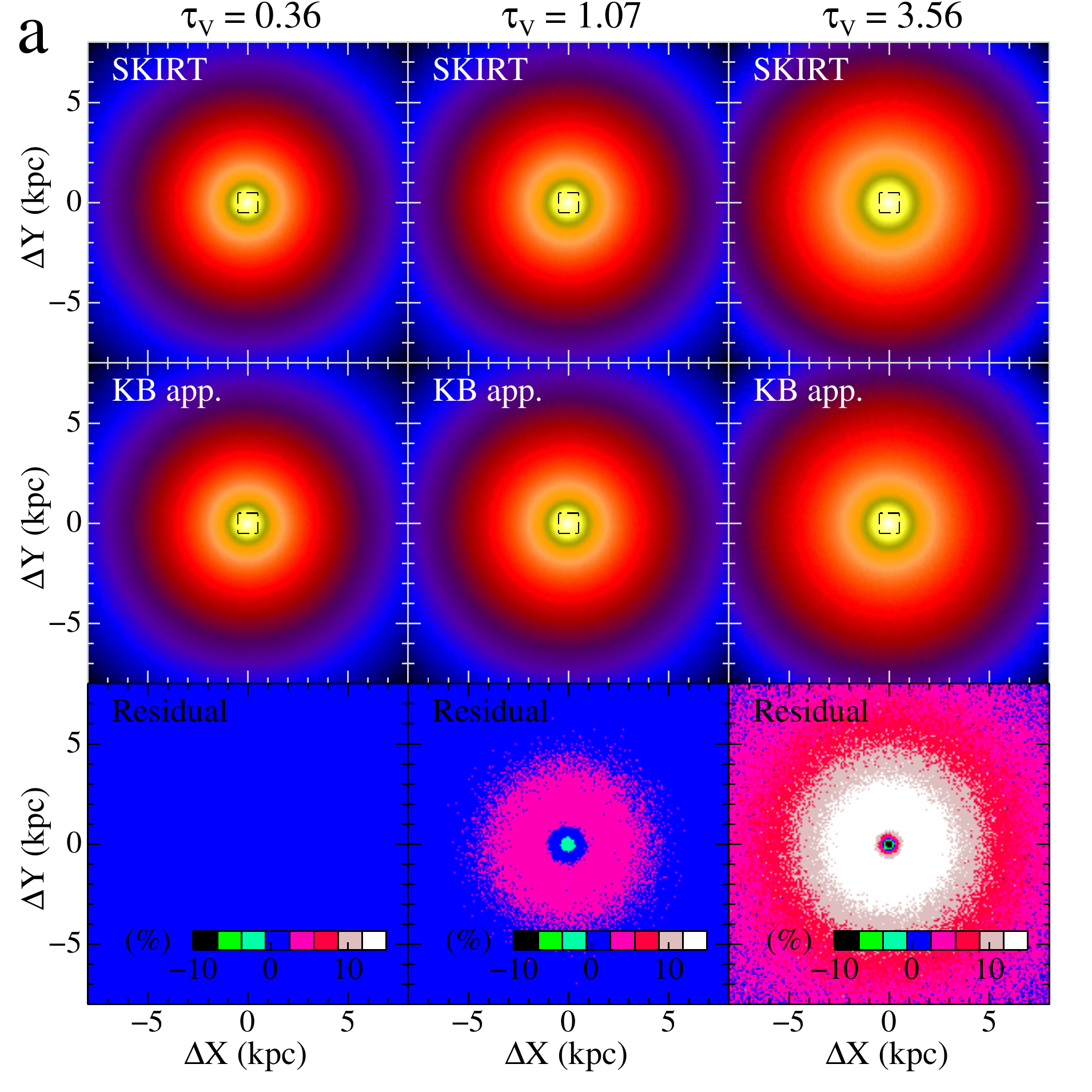}%
\includegraphics[width=0.5\textwidth]{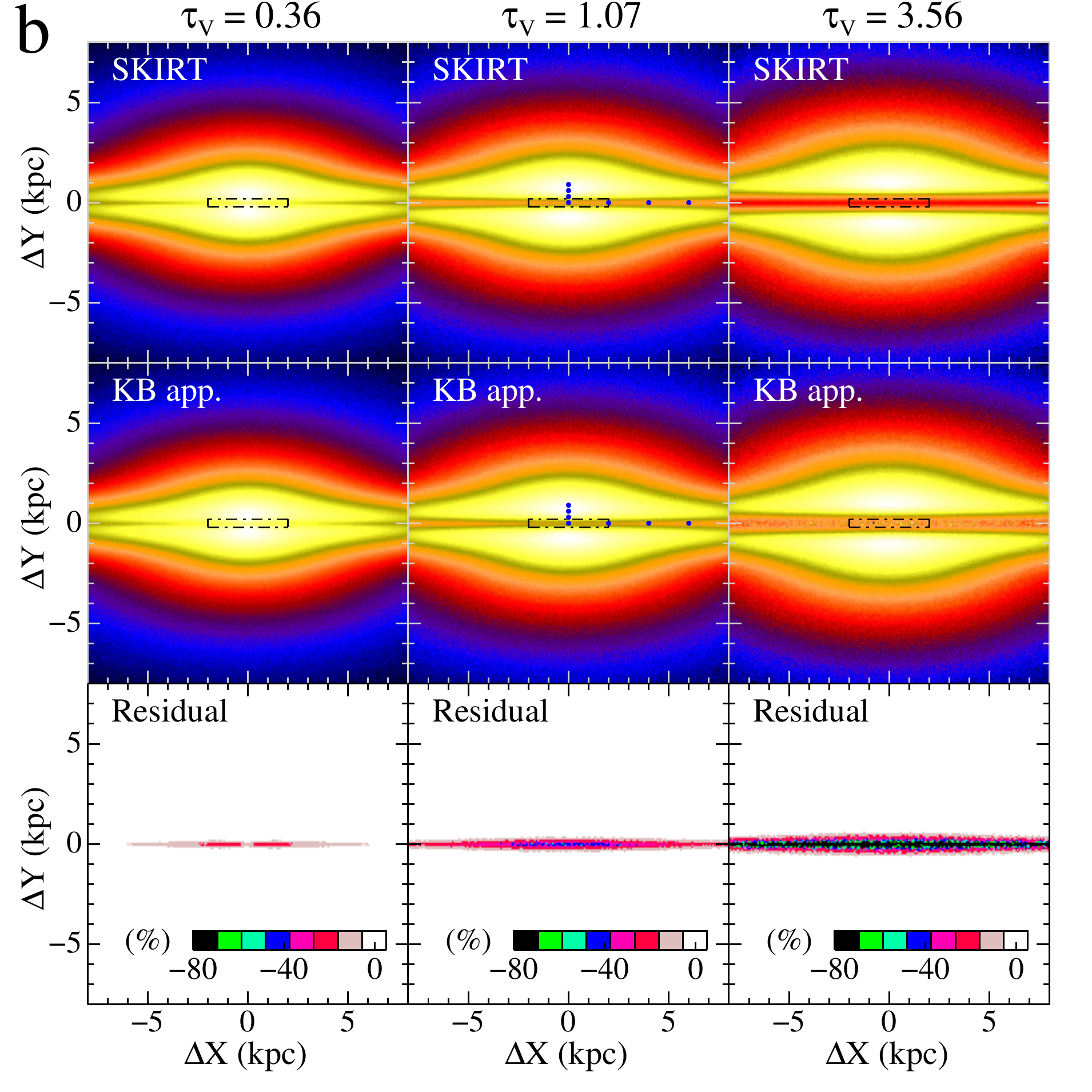}
\vspace*{-1.5em}
\caption{Result images obtained from the \textsc{skirt} code without the KB approximation (top row) and the KB approximation (middle row) for (a) face- and (b) edge-on galaxies. The calculations were performed in the optical V band. The bottom row shows the errors of the KB approximation models, defined by equation (\ref{residual}). Central face-on optical depths are 0.36 (left columns), 1.07 (middle columns) and 3.56 (right columns). The dash-dotted rectangles indicate the central area where the median absolute errors in Table~\ref{minmaxTable} are calculated. Note that, for a better comparison, the contour levels in the residual maps for the edge-on case have been adjusted to have the same range which is smaller than that from the $\tau_{\text{V}} = 3.56$ model.} 
\label{FaceOnEdgeOn.fig}
\end{figure*}

In order to illustrate and explain the main trends, we focus on the face-on and edge-on orientation (Fig.~{\ref{FaceOnEdgeOn.fig}}). The left-most column clearly shows that the KB approximation works very well for the most optically thin galaxy model: the difference between the approximate and full radiative transfer models never exceeds the 1\% level. The discrepancies gradually become large as the optical depth in the model increases. For the $\tau=1.07$ model, the errors are only a few percent, but they grow up to 10 percent or more for the $\tau=3.56$ model. The residual maps show a typical radial pattern: the approximation overestimates the surface brightness in the central bulge-dominated region (i.e. negative values of $\cal{R}$), and underestimates it in the disc-dominated outer regions. In general, we can conclude that the approximation works well in the face-on case for the typical optical depths we expect in face-on galaxies \citep[$\tau_{\text{V}}\sim0.5-1$:][]{1999A&A...344..868X, 2007A&A...471..765B, 2014MNRAS.441..869D}.

The situation is quite different for the edge-on cases in which the optical depth along the galactic plane is strongly enhanced due to projection effects (the edge-on optical depth of the model defined in Table~\ref{ParametersTable} is easily obtained using $\tau^{\text{e}} = h_{R,\text{d}}/h_{z,\text{d}}\tau^{\text{f}} \simeq 26.2 \tau^{\text{f}}$) and the relative star-dust distribution varies strongly depending on the line of sight. While the KB approximation is generally satisfactory at lines of sight above and below the mid-plane, the approximate solution systematically overestimates the surface brightness (or underestimates the extinction) along the dust lane. The accuracy of the approximation, as expected, decreases when the optical depth of the galaxy model increases. Even for the most optically thin model, the discrepancy goes up to 20\%, and it increases to 60\% for the $\tau=1.07$ model and even to a factor of about 70 for the $\tau=3.56$ model (Table~\ref{minmaxTable}). In this latter case, the KB formula~(\ref{Iapprox}) hence turns out to be a poor approximation for the true surface brightness.

\begin{table*}
\centering
\caption{Maximum and median absolute errors of the KB approximation for the models in Fig.~\ref{FaceOnEdgeOn.fig}. The median error is calculated at the pixels in dash-dotted rectangles, shown at the galactic centre in Fig.~{\ref{FaceOnEdgeOn.fig}}.}
\label{minmaxTable}
\begin{tabular}{ccccccc}
        \hline
        \vspace{-2.8 mm}  \\
        Model   & \multicolumn{3}{c}{Face-on galaxy model}      
                                & \multicolumn{3}{c}{Edge-on galaxy model} \\ 
        \vspace{-2.8 mm}  \\
        \begin{tabular}[c]{@{}c@{}} $\tau_{\text{V}}$\end{tabular}
        & \begin{tabular}[c]{@{}c@{}}Maximum\\ negative\\ error (\%)\end{tabular} 
        & \begin{tabular}[c]{@{}c@{}}Maximum\\ positive\\ error (\%)\end{tabular} 
        & \begin{tabular}[c]{@{}c@{}}Median absolute \\ error at the \\ central area (\%)\end{tabular} 
        & \begin{tabular}[c]{@{}c@{}}Maximum \\ negative \\ error (\%)\end{tabular} 
        & \begin{tabular}[c]{@{}c@{}}Maximum\\ positive \\ error (\%)\end{tabular} 
        & \begin{tabular}[c]{@{}c@{}}Median absolute \\ error at the \\ central area (\%)\end{tabular} \\
        \vspace{-2.8 mm}  \\
        \hline
        \vspace{-2.8 mm}  \\
        0.36 & $-0.21$	& 0.95	& 0.09 & $-23.07$		& 0.67 & 10.64 \\
        1.07 & $-1.79$  	& 4.34	& 0.80 & $-59.20$ 		& 2.37 & 27.68 \\
        3.56 & $-10.40$	& 16.23	& 7.88 & $-1704.63$	& 4.15 & 70.17 \\ 
        \vspace{-2.8 mm}  \\
        \hline
\end{tabular}
\end{table*}

\begin{figure*}
\includegraphics[width=0.7\textwidth]{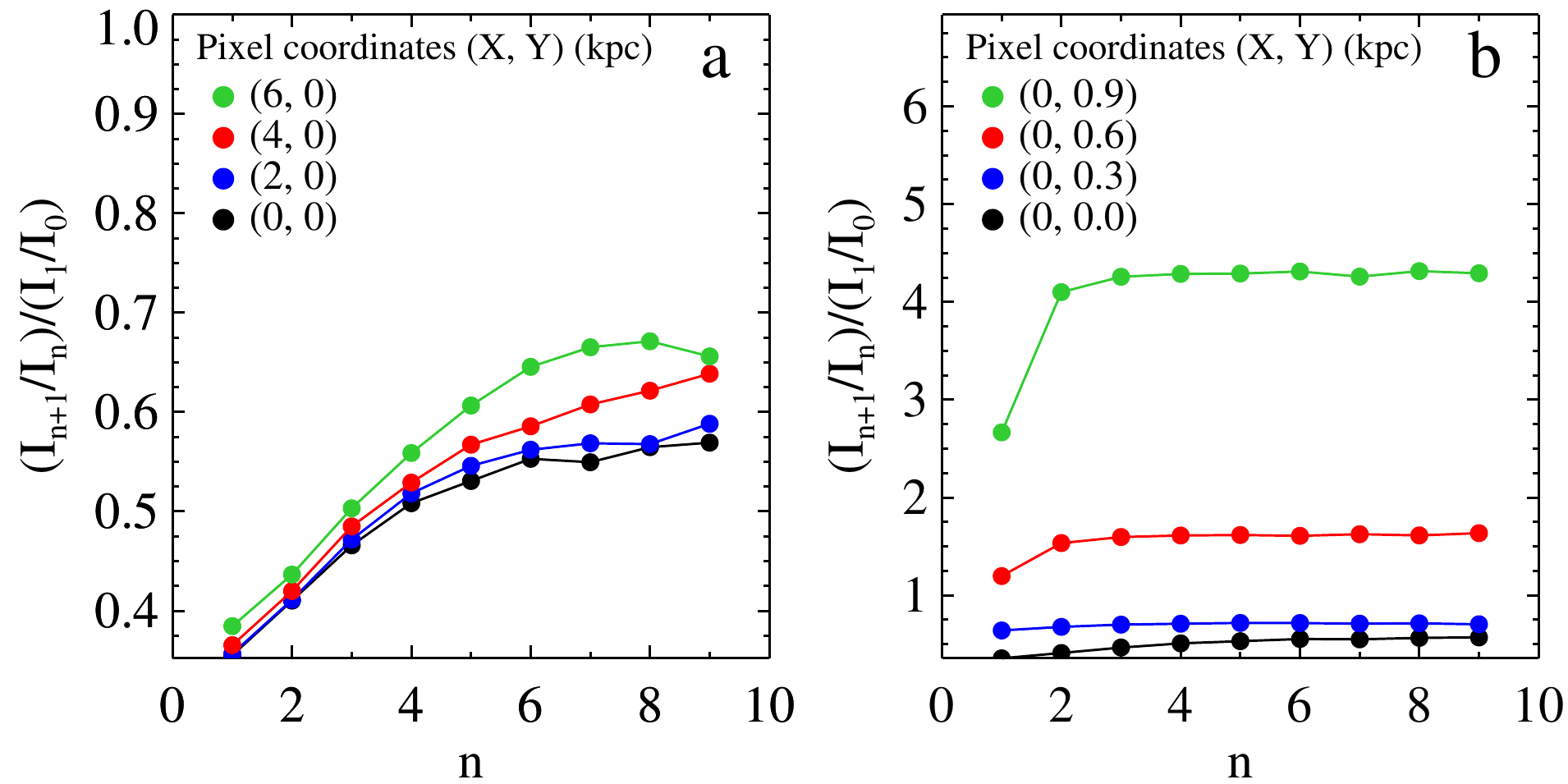}
\caption{Scattering term ratios $S_n$, defined by $(I_{n+1}/I_{n})/(I_1/I_0)$, for several scattering order $n$ measured at various locations denoted as blue dots in Fig.~\ref{FaceOnEdgeOn.fig} for an edge-on galaxy model with $\tau_{\text{V}} = $ 1.07: (a) ratios for three horizontal positions (blue, red and green filled circles) located at intervals of 2 kpc and (b) ratios for three vertical positions located at intervals of 0.3 kpc from the centre. Black filled circles denote the values measured at the galactic centre in both figures. 
}
\label{Ratios.fig}
\end{figure*}

To understand the cause of these large discrepancies, we plot  the ratio
\begin{equation}
S_n \equiv \frac{I_{n+1}/I_n}{I_1/I_0} 
\label{ratio}
\end{equation}
in Fig.~{\ref{Ratios.fig}} as a function of the scattering level $n$ for the edge-on model with $\tau=1.07$. These ratios were derived at seven locations within the image, in order to examine regional variation. The locations are chosen along the major and minor axes of the galaxy image; they are marked as blue dots in the right panel of Fig.~{\ref{FaceOnEdgeOn.fig}} and their coordinates are indicated in Fig.~{\ref{Ratios.fig}}. To reduce Poisson noise, we have used $2 \times 10^{10}$ photon packages, 100 times more than those used in computations of Figs \ref{ModelResults.fig} and \ref{FaceOnEdgeOn.fig}, and calculated median of $3 \times 3$ pixels around the locations. If the assumption (\ref{assumption}) holds, this ratio should be equal to one. However, it is obvious that this is not the case along any line of sight.

At the positions along the major axis (left panel of Fig.~{\ref{Ratios.fig}}), the ratio increases with the value of $n$, but the ratio consistently remains below 1. Consequently, the KB approximation overestimates each term in the series~(\ref{I}), which leads to a significant underestimation of the extinction. This is caused by a complex interplay of absorption, scattering and the star-dust geometry. In our model, the dust distribution is more radially extended than the stellar distribution and the optical depth near the galactic center is much larger than 1. Thus, the direct intensity $I_0$ is strongly attenuated in the galactic plane and, sometimes, becomes even lower than the singly scattered intensity $I_1$, implying that $I_1/I_0>1$.  It should also be noted that, as photons experience more scattering events, the photons are more radially dispersed. This implies that $I_{n+1}/I_n<1~(n\ge1)$ at the central region. As a result, we obtain $S_n<1$.

The right panel of Fig.~{\ref{Ratios.fig}} shows that the assumption~(\ref{assumption}) does not hold along the minor axis either. The figure shows that $S_n < 1$ at $|z|\lesssim300$ pc, as in lines of sight along the major axis. However, the ratio $S_n$ is larger than 1 at distances higher than 300 pc from the galactic plane; the ratio larger than 4 is found at $|z|=900$ pc. This might seem somewhat surprising, as the optical depth is not large anymore along this line of sight. The explanation for this behaviour is the strong difference between the stellar and dust distributions along these lines of sight. The radiation field at high altitudes is dominated by direct starlight from the bulge and the scattered light plays only a minor role (i.e. $I \approx I_{0}$ at $|z| > 2$ kpc). This yields $I_0 \gg I_1$ and  thus $S_n>1$ at high altitudes. Consequently, the KB approximation underestimates the true surface brightness. However, the underestimation  at high altitudes, in spite of fairly large $S_n$ ratio, is not significant because the direct intensity $I_0$ dominates the surface brightness.

\begin{figure}
\centering
\includegraphics[angle=-90, width=0.75\columnwidth]{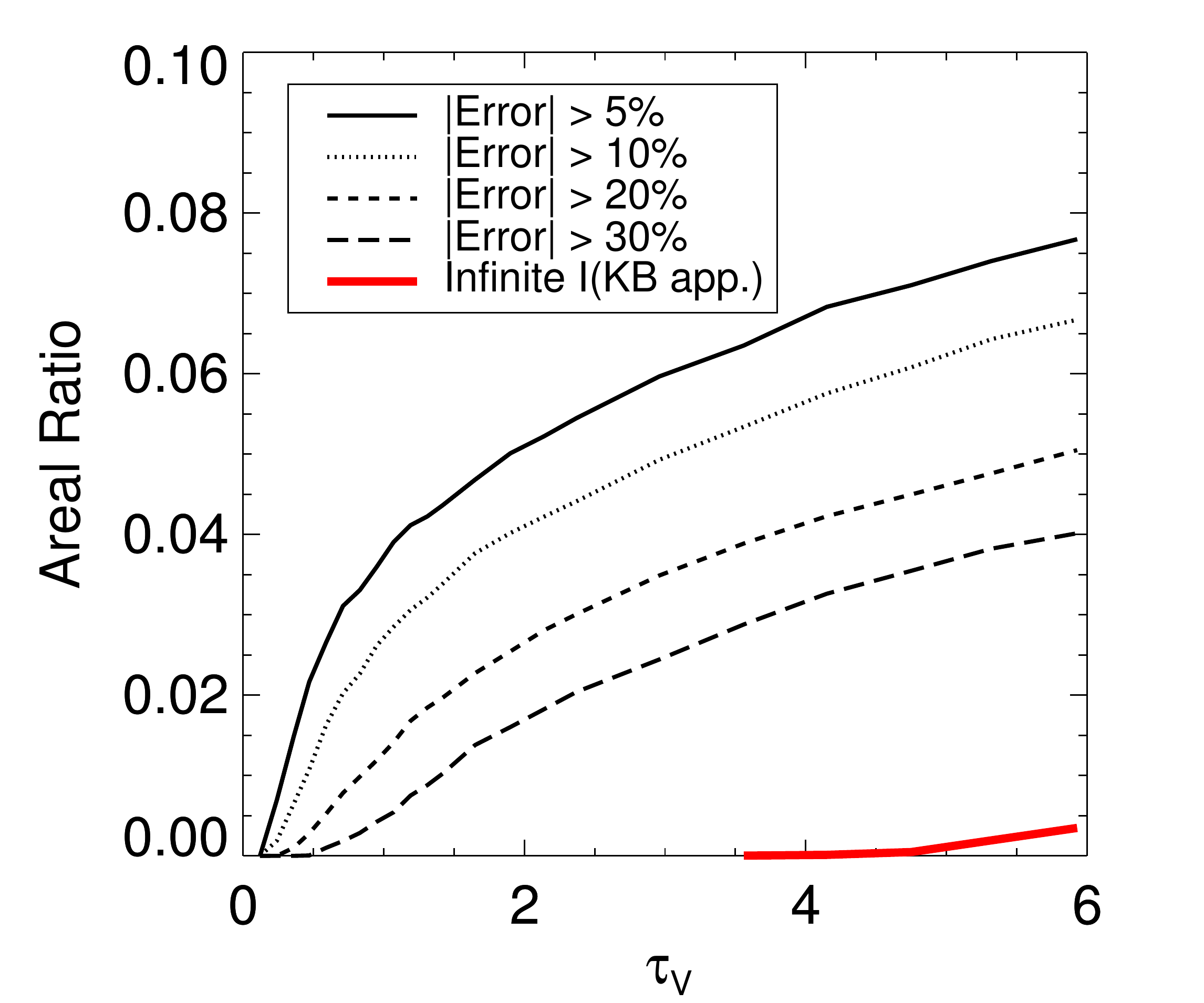}
\caption{Fractions of pixels of which the absolute error of the KB approximation method is larger than a certain minimum value as a function of optical depth. The fractions are calculated for the minimum error of 5\% (solid line), 10\% (dotted line), 20\% (dashed line) and 30\% (the long dashed line). The thick red line shows the fraction of area over which the approximate method yields infinite intensity.}
\label{ArealRatioEdgeOn.fig}
\end{figure}

Fig.~{\ref{ArealRatioEdgeOn.fig}} shows the fraction of pixels in the edge-on galaxy models of which the absolute residual $|\cal{R}|$ is larger than 5, 10, 20, and 30\%. The fraction depends on the size of the image and thus  the value itself is not important. The figure shows a clear correlation between the areal fraction (or the number of pixels) with residuals above a certain threshold and the central face-on optical depth. 

Fig.~{\ref{ArealRatioEdgeOn.fig}} also shows that the KB approximation is not always physical. The approximate equation~(\ref{Iapprox}) is valid only when $I_1<I_0$, i.e. when the contribution of direct light to the surface brightness exceeds the contribution due to singly scattered radiation. This condition can be satisfied at low optical depths ($\tau\lesssim1$) and at wavelengths where the scattering albedo is about 0.5. At some lines of sight with high optical depths ($\tau>1$) around the dust lane, however, the direct starlight was found to be slightly lower than the singly scattered intensity. This can happen when  the dust scalelength is larger than the stellar scalelength and/or the central face-on optical depth is large enough. In this case, the KB approximation is inapplicable; applying equation~(\ref{Iapprox}) results in negative intensities and equation~(\ref{Iwithassumption}) diverges to infinity. The red curve in Fig.~{\ref{ArealRatioEdgeOn.fig}} indicates the number of pixels in which the approximation simply breaks down. This starts to happen at $\tau\sim4$,  and the fraction of pixels with infinite intensity increases with increasing central face-on optical depth.

\begin{figure*}
\centering
\includegraphics[width=0.5\textwidth]{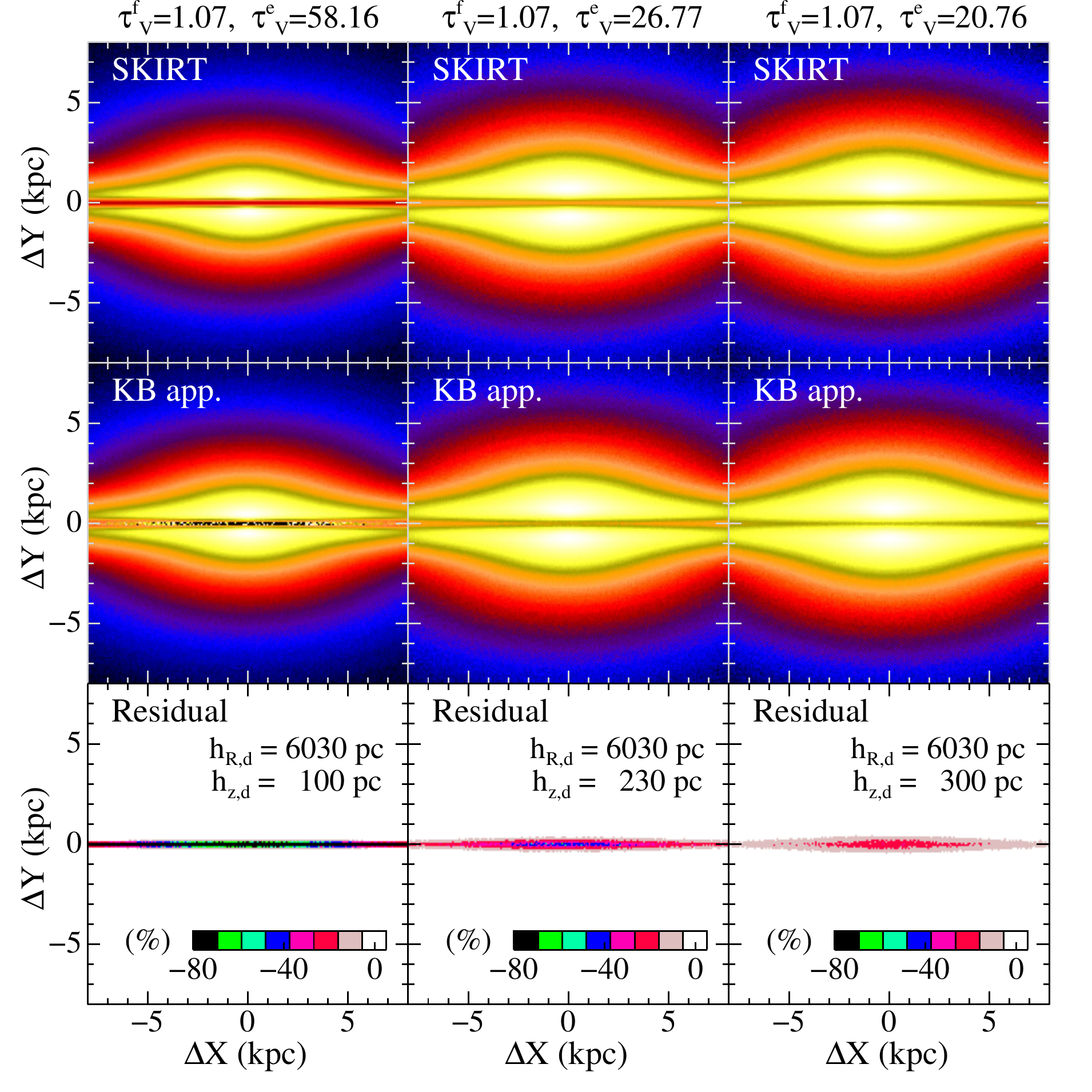}%
\includegraphics[width=0.5\textwidth]{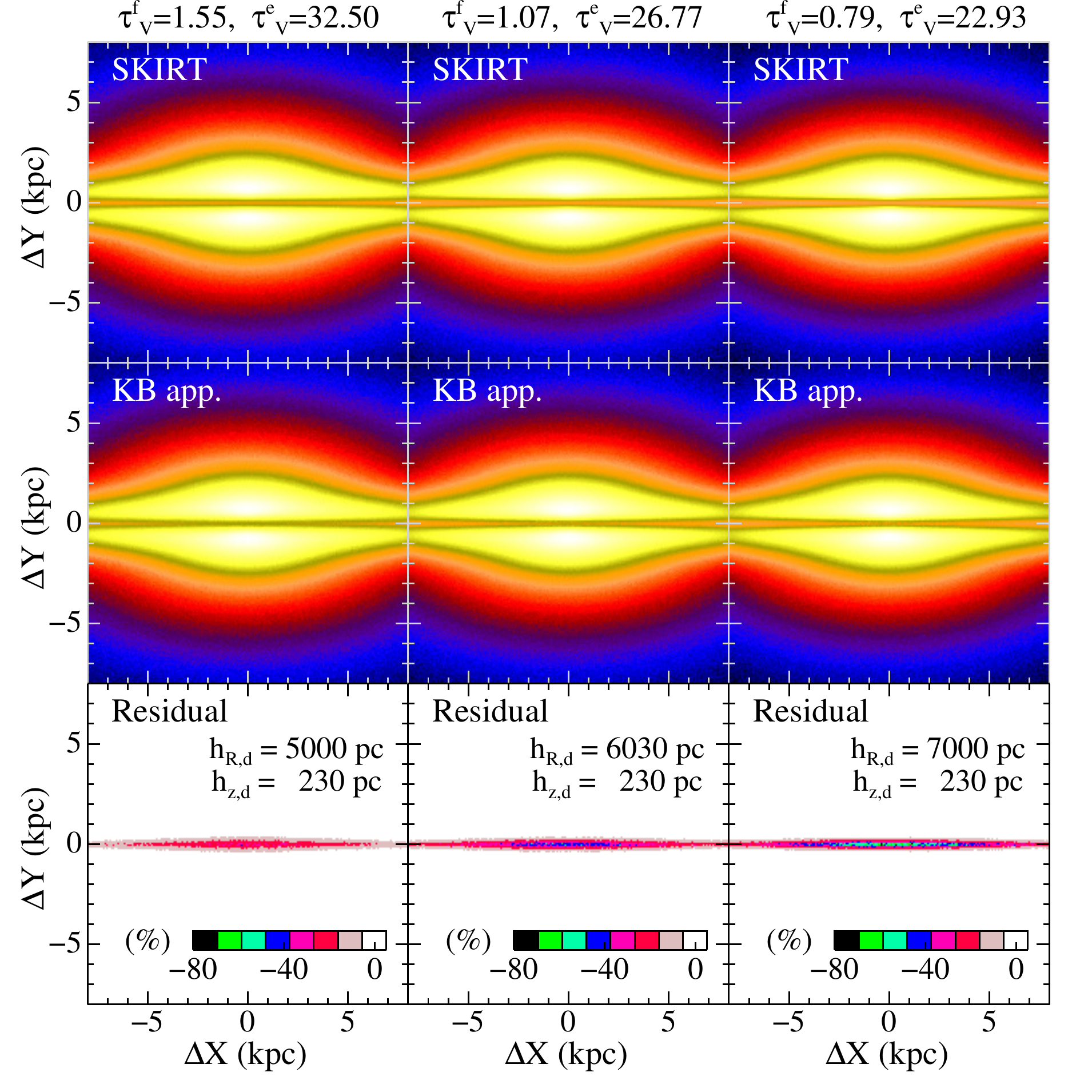}
\vspace*{-1.5em}
\caption{Result images of the \textsc{skirt} code (top row) without the KB approximation, the KB approximation (middle row) and the error (bottom row) defined by equation (\ref{residual}) for edge-on galaxy models with various dust scalelengths ($h_{R,\text{d}}$) and scaleheights ($h_{z,\text{d}}$), but with the same total dust mass and stellar distribution. The calculations were performed in the optical V band. Resultant central face-on ($\tau^{\text{f}}_{\text{V}}$) and edge-on optical depths ($\tau^{\text{e}}_{\text{V}}$), observed from the models, are denoted at the top of each column.}
\label{Nonlinear.fig}
\end{figure*}

Last but not least, we note that the error of the KB approximation is nonlinear. It is obvious that the central optical depth and inclination are not the only factors that affect the accuracy of the KB approximation. The strength and spatial variation of the residuals are coupled in a complex way with other model parameters, such as bulge-to-disc luminosity ratio, and the scalelengths and scaleheights of both stars and dust. Apart from the ``standard'' models discussed in Section~{\ref{GalaxyModel.sec}} and presented in Fig.~{\ref{ModelResults.fig}}, we have computed several other galaxy models in which we varied these model parameters. These tests show that, indeed, other parameters also do play a significant role in shaping both the strength and the extent of the residual maps. Fig.~{\ref{Nonlinear.fig}} compares simulated images of edge-on galaxy models with different scaleheights and scalelengths for the exponential dust disc, while keeping the stellar distribution and the total dust mass fixed. In the left-hand panels, only the dust scaleheight is varied; clearly the strength and the spatial variation of the residuals vary significantly, even though the total dust mass and face-on optical depth are the same. In general, the error increases with decreasing dust scaleheight. In the right-hand panels, only the dust scalelength is varied; they show that the magnitude of the residuals can increase as the dust scalelength increases even if the optical depth decreases.

The bottom line is that, due to the complex and nonlinear nature of radiative transfer in a disc galaxy environment, it is essentially impossible to predict the strength of the deviations or the morphology of the residual maps without a direct comparison with the exact solution. In other words, there are no general recipes that can be used to correct for the use of the KB approximation.

\section{Discussion}
\label{Discussion.sec}

\subsection{The accuracy of approximate radiative transfer}

The present study was motivated by our desire to investigate the reliability of the KB radiative transfer approximation proposed by \citet{1987ApJ...317..637K}. This approximation enabled them and many radiative transfer modellers after them to perform calculations which would otherwise have been too expensive. Indeed, the advantage of the approximation is not to be underestimated: the method only requires the calculation and storage of the direct emission and the singly scattered radiation, instead of the entire sequence of high-order scattering terms. As a result, it saves enormously in computing time and memory consumption. 

Today, almost three decades after the introduction of the approximation, computer speed and memory have increased to the degree that full 3D radiative transfer calculations without explicit approximations are possible \citep{2013ARA&A..51...63S}. This enables us to critically test the accuracy of the KB approximation. On the one hand, this is useful to assess the reliability of the method, and the many scientific results that have been obtained with radiative transfer calculations that make use of this approach. On the other hand, the possible applicability of this approximation remains interesting, as a fast and memory-cheap calculation of the surface brightness distribution of dusty galaxies is still important. This is particularly the case in the frame of inverse radiative transfer modelling, in which a large set of radiative transfer models are calculated to fit observational data \citep{2007A&A...471..765B, 2012ApJ...746...70S, 2013A&A...550A..74D, 2014MNRAS.441..869D, 2016arXiv160506239M}.

In the previous section we have found that the approximation works very well in the case of face-on galaxies in the low optical depth regime. This is not because approximation (\ref{Ratios.fig}) is satisfied, but rather because the contribution of scattered light is negligible compared to the direct intensity $I_0$. Effectively, the KB approximation agrees well with the full radiative transfer solution even though the approximation (\ref{Ratios.fig}) is not valid. But as soon as the optical depth or inclination increase, the residuals between the true and approximate solutions grow larger, and purely edge-on models (especially near plane) are affected most. Note that even for the edge-on models, the KB approximation results at the optically thin lines of sight above the mid-plane give fairly good agreement with those of the full radiative transfer calculation. It should also be noted that sometimes the approximate solution is even inapplicable, because the intensity due to singly scattered radiation exceeds the direct intensity.

\subsection{Comparison with previous work}
\label{PreviousWork.sec}

Our work is not the first one that has investigated the accuracy of the KB approximate radiative transfer algorithm: there have been efforts among the users of the KB approximation to verify its accuracy. In fact, \citet{1987ApJ...317..637K} attempted to quantify the reliability of their approximation themselves. Using a galaxy model similar to ours, they explicitly calculated the terms $I_0$, $I_1$ and $I_2$ in the expansion~(\ref{I}), and found that the assumption~(\ref{assumption}) was accurate to within 30\% for $n=1$. They subsequently considered it reasonable to assume that the assumption would also hold for $n>1$, but due to computational limitations they were unable to test that claim any further. 

\citet{1997A&A...325..135X, 1998A&A...331..894X, 1999A&A...344..868X} used the approximate radiative transfer approach in their landmark series of radiative transfer models for a set of edge-on spiral galaxies. In the first paper of this series, they reported that the error in the intensity introduced by the approximate method is typically less than 1\%, but they do not provide further details on how this estimate was achieved. 

A very interesting test of the KB approximation was performed by \citet{2002A&A...384..866M}. They used a Monte Carlo radiative transfer code \citep{1996ApJ...465..127B, 2000MNRAS.311..601B} to generate images of edge-on spiral galaxy models, and subsequently fitted them using the approximate KB method. They found very good agreement: for low optical depths, the fitted parameters were found to be very close to the input values, and even for a face-on optical depth $\tau=5$, the derived optical depth obtained from the fitting was only 15\% larger than the input model. This seems at odds with our results, as we find significant deviations for the edge-on models, and even enter the regime where the approximation breaks down at $\tau\gtrsim4$. 

It is hard to believe that the differences between \citet{2002A&A...384..866M} and our results could be due to the different Monte Carlo codes used. Indeed, \textsc{skirt} and \textsc{trading} \citep{2008A&A...490..461B}, the modern version of the \citet{1996ApJ...465..127B, 2000MNRAS.311..601B} code, are currently involved in a detailed 3D dust radiative transfer benchmarking effort, and their results are found to be compatible to the 1\% accuracy level \citep{2015A&A...580A..87C, Gordon2016}. 

On the contrary, we argue that the discrepancy is mainly caused by differences in the star-dust geometry between both models. The model of \citet{2002A&A...384..866M} contains only two components (a stellar disc and a dust disc), whereas our model contains an additional stellar bulge. Moreover, they truncate the dust disc at 4 vertical scalelength and 4.6 radial scaleheight, whereas the dust disc in our model is essentially untruncated (it extends to 60 kpc and 20 kpc in the radial and vertical directions, respectively). Most importantly, they assume the same scalelengths for the dust and stellar discs. In our model, however, the dust scalelength is 50\% larger than the stellar scalelength, as typically seen in real edge-on spiral galaxies \citep[e.g.][]{1999A&A...344..868X, 2007A&A...471..765B, 2014MNRAS.441..869D}.

We have run a number of additional simulations to investigate the effect of these differences in model setup. When we use an identical setup as \citet{2002A&A...384..866M}, we recover percent-level agreement between the full solution and the KB approximation, for edge-on models and even for optical depths $\tau_{\text{V}}>1$. For example, we find a maximum negative error of --3.3\% for the KB approximation in the edge-on $\tau_{\text{V}}=4$ model. When we change the dust to stellar scalelength ratio to 1.5, this value increases by a factor 6. If a stellar bulge is added and the dust disc truncation is lifted, the value increases again with an additional factor 5--6. These results imply that the reliability of the KB approximation is strongly affected by the adopted stellar and dust distributions; the error becomes more significant as more stars are enclosed within the dust distribution. The star-dust geometry is also expected to affect the ratio of the singly scattered to direct intensity, as discussed in Section~{\ref{Results.sec}}.

\subsection{Implications for radiative transfer modelling}

Our modelling indicates that the KB approximation is accurate for low optical depth and face-on orientations, but that the errors due to the approximation are not negligible at high inclinations. Especially for edge-on orientations the difference between the full solution and the KB approximation can be substantial, even for relatively low optical depths ($\tau<1$). Ironically, radiative transfer modelling studies of galaxies have almost exclusively focused on edge-on spiral galaxies, because the dust obscures a large fraction of the emitted starlight in this configuration, and the projection smears out many of the details and structures that complicate the modelling at smaller inclinations \citep{2000A&A...353..117M, 2002A&A...384..866M, 2015A&A...576A..31S}.

The residuals between the full method and the KB approximation are highest around the dust lane, where the optical depth along the line of sight is largest. It can therefore be expected that, when radiative transfer models are fitted to observed images, some galaxy parameters are less strongly affected than others. For instance, the parameters corresponding to the stellar disc and bulge are probably less sensitive to the use of the KB approximation, because their value is mainly determined from the parts of the image above and below the dust lane. For the actual parameters of the dust distribution, which is usually the goal of the radiative transfer modelling, this is not the case: they are mainly determined from fitting the dust lane, and the outskirts of the galaxy are less sensitive to their value. Our results hence suggest that the KB approximation should be applied with care when it is used to determine the properties of the dust distribution in edge-on spiral galaxies.

Particularly interesting is the difference in the ``depth'' of the dust lane between approximate and full method: for a given optical depth or dust mass, the KB approximation underestimates the extinction in the dust lane (right panel in Fig.~{\ref{FaceOnEdgeOn.fig}}). When observations of real edge-on spiral galaxies are modelled using the KB approximation, one might hence expect that the inferred dust mass would be overestimated. This is extremely relevant in the frame of the discussion around the so-called dust energy balance, the apparent inconsistency between the dust seen in absorption at optical wavelengths, and the dust mass measured in emission from far-infrared observations \citep{2000A&A...362..138P, 2005A&A...437..447D, 2010A&A...518L..39B, 2012MNRAS.427.2797D, 2015MNRAS.451.1728D}.

Whether or not the use of the KB approximation plays a significant role, is hard to predict. Indeed, the importance of the KB approximation depends in a nonlinear way on many other parameters, and isolating the effect of the KB approximation from other systematic differences is hard (see Sections~{\ref{Results.sec}} and {\ref{PreviousWork.sec}}). One way to test this conjecture would be to model identical data using exactly the same model, but just with the KB approximation switched on and off. This falls outside the scope of this paper, but would be an interesting future project.

\begin{figure*}
\centering
\includegraphics[width=0.95\textwidth, trim={0 0 0 0}, clip]{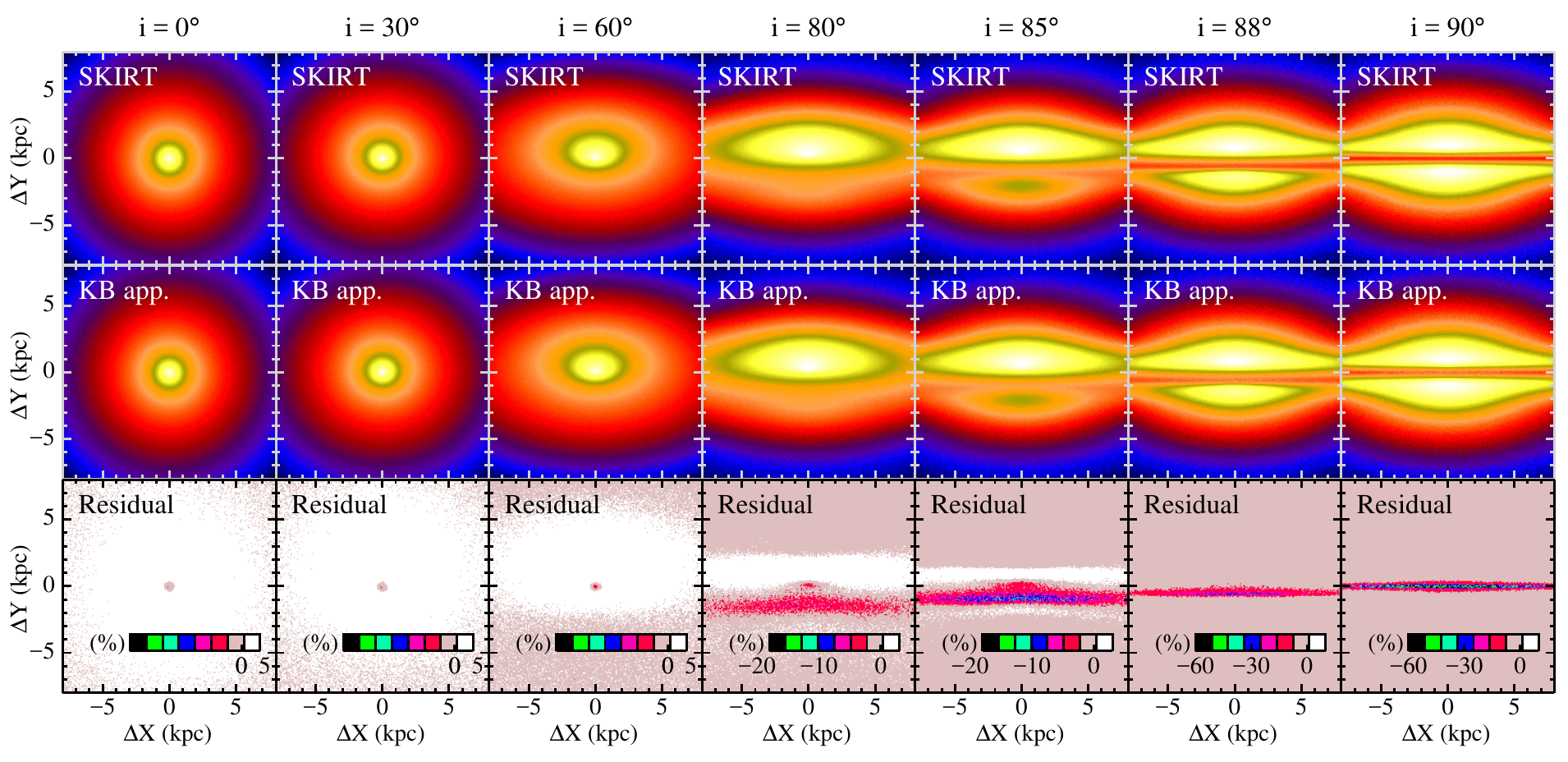}
\caption{Same as Fig.~{\ref{ModelResults.fig}}, but for the central wavelength of the FUV band (1538.6 \AA) and $\tau_{V}$ = 1.07. In the case of the edge-on model ($i = 90\degr$), the median absolute error at the central region and the maximum negative error are about 40\% and $-148\%$, respectively.} 
\label{FUV_Results.fig}
\end{figure*}

Last but not least: we have seen that the effects of the KB approximation are still reasonable for the optical depth values typically encountered at optical wavelengths ($\tau_{\text{V}}\approx0.5-1$), but that the discrepancy gets significantly worse when the optical depth increases. There have been claims in the literature that the typical face-on V-band optical depth is actually larger than 1 \citep[e.g.,][]{2007MNRAS.379.1022D, 2016MNRAS.459.2054D}. But even in galaxies with moderate optical depths at optical wavelengths, this issue becomes relevant when moving to shorter wavelengths. In particular, the FUV is an important wavelength regime for attenuation and radiative transfer studies, because of its importance for estimating star formation rates \citep{2007ApJS..173..267S, 2012ARA&A..50..531K} and the UV slope \citep{1996A&A...306...61B, 2011MNRAS.415.1002W, 2016A&A...586A..13V}. Moreover, the FUV offers important diagnostics about the nature of the dust itself, since it has been shown that the UV extinction curve depends strongly on environment \citep{1989ApJ...345..245C, 2003ApJ...594..279G, 2015ApJ...815...14C}. If radiative transfer studies are performed in the UV wavebands, approximate methods could produce significant errors even for moderate face-on optical depths ($\tau_{V} \approx 1$). Fig.~{\ref{FUV_Results.fig}} shows the results of the galaxy model ($\tau_{V} = 1.07$) in the FUV band for different inclinations. We find that the median absolute error at the central region is about 40\% and the maximum negative error is about $-148\%$ for the edge-on case ($i = 90\degr$). UV radiative transfer studies of edge-on spiral galaxies \citep[e.g.,][]{2014ApJ...785L..18S, 2015ApJ...815..133S, 2016A&A...587A..86B} should therefore be extremely careful in making simplifying assumptions such as the KB approximation.

\section{Conclusions}
We have critically assessed the strength, weaknesses and validity of the KB approximation by comparing it with the full radiative transfer without the approximation. We find that the KB approximation, which is cheaper and faster than the full radiative transfer, is very accurate in the case of an optically thin and face-on galaxy. 

However, results of our galaxy models show that the KB approximation can produce substantial errors as the optical depth and the inclination angle of the galaxy increase. In general, these errors of the approximate method may result in an overestimation of dust mass, which is relevant to the dust energy balance problem. We also find that the KB approximation for our galaxy models yields $I_{1} > I_{0}$ along some lines of sight and thus produces infinite intensities. The inapplicability and significant errors of the KB approximation are mainly found around the dust lane, where the optical depth is highest. This suggests that one should be careful especially in modelling an edge-on spiral galaxy with the KB approximation, because the parameters of the dust distribution are mainly determined from fitting the dust lane. More importantly, it is hard to predict a general trend to correct the error of the KB approximation for various types of galaxy models, due to its nonlinear character. Therefore, it is recommended to use full Monte Carlo radiative transfer without the KB approximation to avoid all the aforementioned errors for the radiative transfer study of spiral galaxies.

\section*{Acknowledgements}

This research was supported through the Interuniversity Attraction Poles Programme initiated by the Belgian Science Policy Office (AP P7/08 CHARM), and through DustPedia, a collaborative focused research project supported by the European Union under the Seventh Framework Programme (2007-2013) call (proposal No. 606824). The participating institutions are: Cardiff University, UK; National Observatory of Athens, Greece; Ghent University, Belgium; Universit\'{e} Paris Sud, France; National Institute for Astrophysics, Italy; and CEA (Paris), France. WH thanks for the support by the National Research Foundation of Korea with grant number NRF-2014M1A3A3A02034746.



\bibliographystyle{mnras}
\bibliography{References}






\bsp    
\label{lastpage}
\end{document}